\documentclass[aps,prb,graphicx,preprint]{revtex4-2}

\usepackage{hyperref}
\usepackage{amsmath}
\usepackage{array}
\usepackage{dcolumn}
\usepackage{graphicx}
\usepackage{grffile}
\usepackage{siunitx}
\usepackage{mw}
\usepackage{bm}
\usepackage{multirow}
\usepackage{tabularray}
\usepackage[table,xcdraw]{xcolor}
\usepackage[outdir=./]{epstopdf}
\begin{document}

\title{\textit{Ab-initio} tensile tests applied to BCC refractory alloys}

\author{Vishnu Raghuraman}
\affiliation{Department of Physics, Carnegie Mellon University, Pittsburgh, PA 15213}

\author{Saro San}
\affiliation{NETL Support Contractor, Albany OR 97321}

\author{Michael C. Gao}
\affiliation{National Energy Technology Laboratory, Albany OR 97321}

\author{Michael Widom}
\affiliation{Department of Physics, Carnegie Mellon University, Pittsburgh, PA 15213}

\begin{abstract}
Refractory metals exhibit high strength at high temperature, but often lack ductility. Multiprinciple element alloys such as high entropy alloys offer the potential to improve ductility while maintaining strength, but we don't know \textit{a-priori} what compositions will be suitable. A number of measures have been proposed to predict the ductility of metals, notably the Pugh ratio, the Rice-Thomson D-parameter, among others. Here we examine direct \textit{ab-initio} simulation of deformation under tensile strain, and we apply this to a variety of Nb- and Mo-based binary alloys and to several quaternary alloy systems. Our results exhibit peak stresses for elastic deformation, beyond which defects such as lattice slip, stacking faults, transformation, and twinning, relieve the stress. The peak stress grows strongly with increasing valence electron count. Correlations are examined among several physical properties, including the above-mentioned ductility parameters.

\end{abstract}
\maketitle

\section{Introduction}

There is an ever-growing demand for alloys with high strength and ductility. Refractory high entropy alloys are potential candidates \cite{rhea-features-1,rhea-features-2}, with the added advantage of high melting points \cite{rhea-features-3,rhea-features-4}, but identifying suitable alloys among an extremely large sample space can be a challenge. First principles approaches to alloy design offer a potentially simple and efficient way to screen for promising alloy compositions \cite{first-principles-1,first-principles-2}, which can then be synthesized and tested experimentally. While some mechanical properties like bulk and shear modulus can be accurately determined by DFT \cite{elastic-constants-1,elastic-constants-2,elastic-constants-3}, ductility is much harder to predict \cite{ductility-review} because distinct physical mechanisms enhance or diminish ductility, making it challenging to develop a single numerical parameter or test that is material agnostic. Additionally, extrinsic effects such as defects or impurities may mask the intrinsic ductility of a bulk single phase.

Experimentally, ductility can be estimated by performing a tensile test, where a sample of the alloy is subjected to uniaxial stress. Initially, the system deforms elastically - it returns to its undeformed shape on the removal of the applied uniaxial stress. When the applied stress exceeds the yield strength, the material deforms permanently. This is referred to as plastic deformation. On further application of stress, the system continues to deform plastically until it breaks. Ductility can be described as the ability of the system to undergo plastic deformation before tensile failure \cite{ductility-definition}. An alloy with a larger plastic region is considered to be more ductile. 


The onset of plastic deformation can be triggered by several types of defects. A portion of the crystal may shear with respect to the other along a particular plane. This can be caused by the passage of a dislocation that restores the crystalline order after lattice planes slip with respect to each other. The dislocation displacement and magnitude is represented by the Burgers vector. Plasticity may also be facilitated by stacking faults, which are errors in the arrangement of atomic planes and can be caused by the passage of partial dislocations \cite{dislocations-1,dislocations-2}. Additionally, certain atoms may shear to produce a new equivalent crystal grain that shares a boundary with the original grain but at a different orientation. This process, referred to as twinning \cite{twinning}, can contribute to the ductility of the system, as observed in Twinning-Induced Plasticity (TWIP) steel \cite{twip-steel}. Alternatively, the crystal may distort into an entirely new crystallographic structure, leading to transformation-induced plasticity (TRIP) \cite{trip-1,trip-2}. Dislocation slip, deformation, and twinning compete, and depending on the system, any or all may facilitate plasticity.

A number of dimensionless measures have been proposed to predict the ductility of metals, notably the Pugh ratio of shear over bulk modulus~\cite{pugh}, and the Rice-Thomson D-parameter given by the ratio of unstable stacking fault energy to surface energy~\cite{rice,rice-thomson,tb-rice,zu-curtin,tu-gsfe}. Another dimensionless parameter, $\chi$, related to the strain thresholds for mechanical instabilities was recently proposed~\cite{winter,dejong}. Each of these can be readily calculated \textit{ab-initio}, yet none proves fully satisfactory. Here we evaluate the utility of an \textit{ab-initio} tensile test in which we relax structures under increasing tensile strains and follow their mechanical stability and deformation. This approach can supplement the information obtained from the dimensionless parameters. In particular, it reveals peak stresses beyond which lattice slip, stacking faults, structural transformation, twinning, and other deformations, relieve the stress, potentially aiding in ductility.

We apply this computational technique to study refractory Mo- and Nb-based BCC binary and quaternary alloys under tension along the [001] direction. Along with the peak stress, we calculate the valence electron count, the Pugh ratio, and the D and $\chi$ parameter. In addition we calculate the Fermi level electronic density of states, which reveals the balance of metallic vs. covalent bonding, and the eigenvalues of the elasticity tensor, which reveal the mechanical stability of the structure. These quantities are shown to be correlated among each other. For example, high valence electron count is associated with low density of states, D parameter and Pugh ratio, which in turn are indicative of low ductility. Detailed quantitative information and records of deformation type for each alloy are recorded in the Supplemental Information \cite{supplementary}.


\section{Methodology}
\subsection{Elastic Constants}
Consider a crystalline system at equilibrium, with volume $V_0$. If a Lagrangian strain $\eta$ is imposed on the system, the corresponding elastic energy can be expressed as
\begin{equation}
\Delta E = \frac{1}{2}C_{ij}\eta_i\eta_j + O(\eta^3)
\label{eq:elastic-energy}
\end{equation}
where
\begin{equation}
	C_{ij} = \frac{1}{V_0}\left(\frac{\partial^2 E}{\partial \eta_i \partial \eta_j}\right)_{\eta=0}
\end{equation} 
is the second order elastic constant, expressed in Voigt notation. For a cubic crystal, this has the form
\begin{equation}
	C = \begin{pmatrix}
		C_{11} & C_{12} & C_{12} & 0 & 0 & 0 \\
		C_{12} & C_{11} & C_{12} & 0 & 0 & 0 \\
		C_{12} & C_{12} & C_{11} & 0 & 0 & 0 \\
		0 & 0 & 0 & C_{44} & 0 & 0 \\
		0 & 0 & 0 & 0 & C_{44} & 0 \\
		0 & 0 & 0 & 0 & 0 & C_{44}
		\end{pmatrix}.
\end{equation}
There are only three independent elastic constants - $C_{11}$, $C_{12}$ and $C_{44}$. The bulk modulus
\begin{equation}
	K = \frac{1}{3}\left(C_{11} + 2C_{12}\right)
\end{equation}
and the shear moduli 
\begin{equation}
	G = C_{44}\;\mathrm{or}\;\mu = \frac{1}{2}\left(C_{11} - C_{12}\right),
\end{equation}
depending on the type of shear. Using the bulk and shear moduli, the Pugh ratio $K/G$ \cite{pugh} can be calculated. This ratio is a well-known measure of ductility - a high Pugh ratio is an indicator of a ductile system.
\subsection{Elastic Stability}
A system at equilibrium is elastically stable when the energy increases on application of infinitesimal strain $i.e$ the elastic energy is positive. Using equation \ref{eq:elastic-energy}, we can express the stability criteria as
\begin{equation}
    C_{ij}\eta_i\eta_j > 0\;\mathrm{or}\;\mathbf{\eta}^{T}\mathbf{C}\mathbf{\eta} > 0,
\end{equation}
where $\mathbf{\eta}$ is any infinitesimal strain applied to the system. This condition is satisfied when $\mathbf{C}$ is a positive definite matrix, or equivalently, if all the eigenvalues of the second order elastic constant matrix are positive. This is the Born stability criterion \cite{Born}.

If the crystal is driven out of equilibrium by a finite strain $\mathbf{\beta}$, the stability is determined by the new condition \cite{wallace,morris}
\begin{equation}
    \lambda_{ij}(\mathbf{\beta})\eta_i\eta_j > 0\;\mathrm{or}\;\mathbf{\eta}^{T}\mathbf{\lambda(\beta)}\mathbf{\eta} > 0,
\end{equation}
where $\mathbf{\lambda (\beta)} $ is the symmetrized Wallace tensor at finite strain $\beta$, \cite{wallace,morris}
\begin{equation}
	\label{eq:Wallace}
	\lambda_{klmn} (\beta) = C^{\prime}_{klmn} (\beta) + \frac{1}{2}\left[\tau_{ml}\delta_{kn} + \tau_{km}\delta_{ln} +      \tau_{nl}\delta_{km} + \tau_{kn}\delta_{lm} - \tau_{kl}\delta_{mn} - \tau_{mn}\delta_{kl}\right],
\end{equation}
where $C^{\prime}_{klmn}(\beta)$ is the second order elastic constant , and $\tau$ is the Lagrangian stress tensor for the non-equilibrium crystal. For a cubic system that has been strained along the [001] axis, the symmetric Wallace tensor in Voigt notation is
\begin{equation}
\mathbf{\lambda}(\beta) = 
\begin{pmatrix}
    C^{\prime}_{11} & C^{\prime}_{12} & C^{\prime}_{13} - \frac{\tau}{2} & 0 & 0 & 0 \\
    C^{\prime}_{12} & C^{\prime}_{11} & C^{\prime}_{13} - \frac{\tau}{2} & 0 & 0 & 0 \\
    C^{\prime}_{13} - \frac{\tau}{2} & C^{\prime}_{13} - \frac{\tau}{2} & C^{\prime}_{33} + \tau & 0 & 0 & 0 \\
    0 & 0 & 0 & C^{\prime}_{44} + \frac{\tau}{2} & 0 & 0 \\
    0 & 0 & 0 & 0 & C^{\prime}_{44} + \frac{\tau}{2} & 0 \\
    0 & 0 & 0 & 0 & 0 & C^{\prime}_{66}
\end{pmatrix}.
\end{equation}
This matrix has six eigenvectors and five unique eigenvalues. The system is stable if all the eigenvalues are positive. If one (or more) of the eigenvalues vanishes, then $\beta$ is a critical strain for mechanical failure. The type of failure can be determined based on the eigenvector.
\begin{itemize}
    \item If the critical eigenvalue has a non-zero component along the strain direction (in this case, the component $v_3$ for an eigenvector $v$), it is considered a ``cleavage"\cite{winter} or ``extensional" \cite{splay} instability and potentially signals a fracture of the solid \cite{winter}.
    \item If the critical eigenvalues has no component along the strain direction, it is considered a ``shear" \cite{winter,splay} instability. The shear mode, for some systems, can lend ductility \cite{splay,qi}.
\end{itemize}

\subsection{$\chi$ parameter}
The Wallace tensor at finite strain can be approximated using the second and third order elastic constants at zero strain. This weakly nonlinear approach allows fast  approximate calculation of the eigenvalues at different values of finite strain. Using this approach, we can determine the critical strain $\eta_c$, corresponding to the first ``cleavage" instability and $\eta_s$, corresponding to the first ``shear" instability. The intrinsic ductility parameter \cite{dejong,winter}
\begin{equation}
    \chi = \frac{\eta_c}{\eta_s},
\end{equation}
determines the systems ability to deform via shear instability before fracture. A system is deemed ductile if the shear instability occurs before cleavage ($\chi > 1$), implying that the system deforms plastically before fracture. Conversely, if the cleavage instability occurs before shear ($\chi < 1$), the system is deemed brittle.
\subsection{D-parameter}
The D-parameter proposed by Rice \cite{rice,tb-rice} is another measure of ductility, given by 
\begin{equation}
    D = \frac{\gamma_{s}}{\gamma_{usf}},
\end{equation}
where $\gamma_{usf}$ is the unstable stacking fault energy and $\gamma_{s}$ is the surface energy. In this approach, ductility is thought to arise from the formation of dislocations at a crack tip, blunting the tip and hence reducing the stress. A high D-parameter corresponds to a ductile material while lower D-parameter corresponds to a brittle material. The D-parameter is a reciprocal of the Rice parameter \cite{rice} $\gamma_{usf}/\gamma_{s}$ and is based on the Rice-Thomson parameter \cite{rice-thomson} $Gb/\gamma_{s}$, where $G$ is the shear modulus and $b$ is the Burgers vector. Both $\gamma_{usf}$ and $\gamma_{s}$ can be calculated directly from first principles, or obtained from a surrogate model trained on first-principles data \cite{surrogate}.
\subsection{\textit{Ab-initio} tensile test}
In this fully nonlinear approach, the second order elastic constants and stress at finite strain are directly calculated without any extrapolations from zero strain. This is done by stretching the equilibrium structure and performing first-principles relaxation in order to obtain the structure at finite strain and the induced stress. The relaxed structure is then used to calculate the second order elastic constants at finite strain. This is referred to as the $ab$-$initio$ tensile test. While it is computationally more intensive as compared to the weakly nonlinear $\chi$-parameter approach, it provides higher accuracy and significantly more insight into the behavior of the system. The first-principles relaxation allows the system to develop stacking faults, dislocations and other interesting effects. In the following section, we discuss the application of this method in detail.
\section{Tensile Test Observations}
We calculated eigenvalues, stress and lattice distortion at fixed uniaxial strain along the [001] direction, for Mo$_3$X and Nb$_3$X, where X $\in$ \{Al, Cr, Hf, Nb/Mo, Re, Ru, Si, Ta, Ti, V, W, Zr\}, and a set of refractory alloy based quaternaries. 128-atom SQS structures were used. A maximum strain of 20\% was applied.To gauge the size effect, tensile test was also peformed for a 54-atom SQS of Mo$_3$Nb. No significant differences were found. Full results for the complete set of structures are provided in the supplementary material, along with the computational details \cite{supplementary}. At certain strains, the relaxed structures show interesting defects, which are discussed in the following subsections. 
\subsection{Stacking Faults}
\begin{figure}
    \centering
    \includegraphics[width=0.49\linewidth]{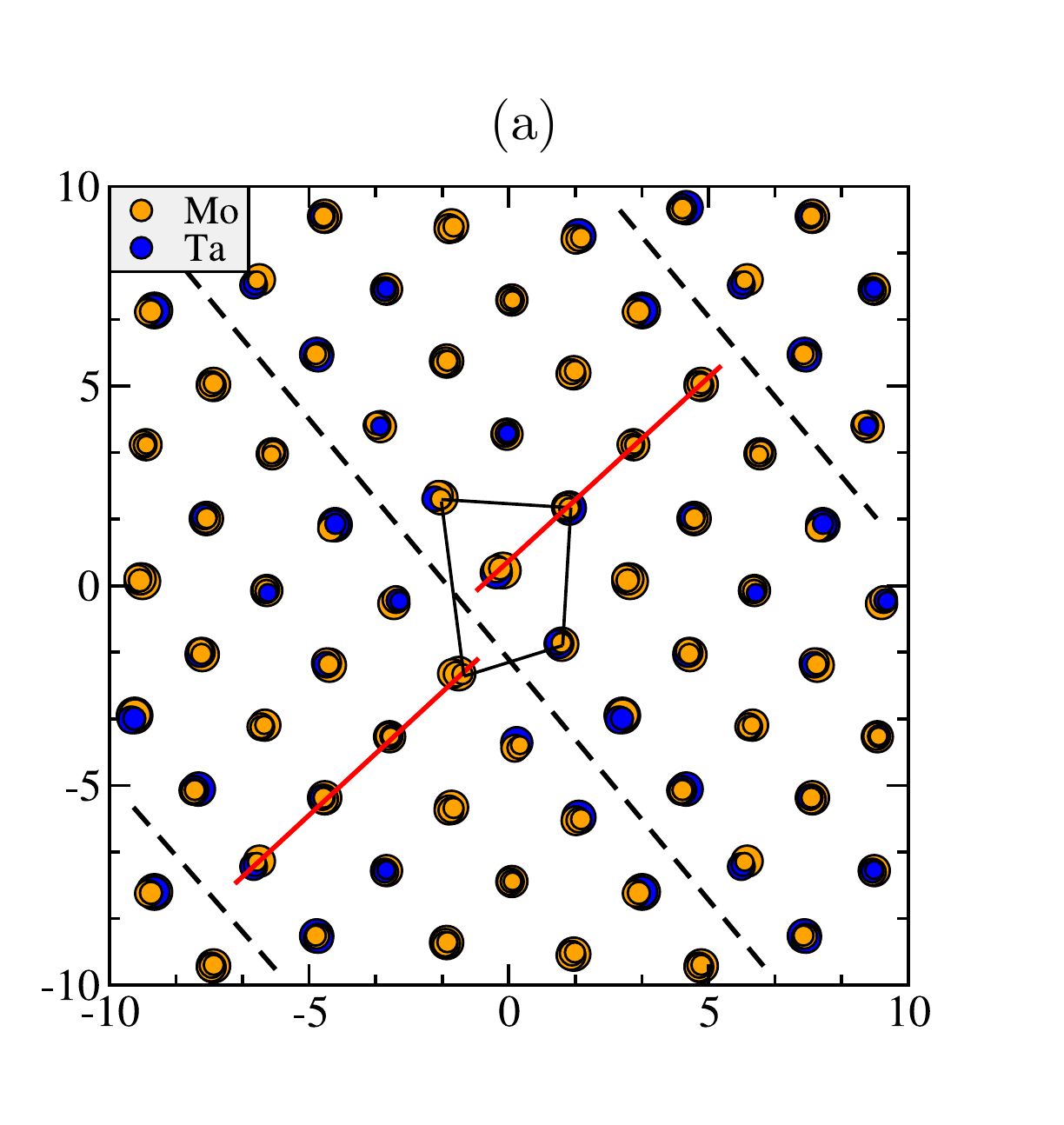}
    \includegraphics[width=0.46\linewidth]{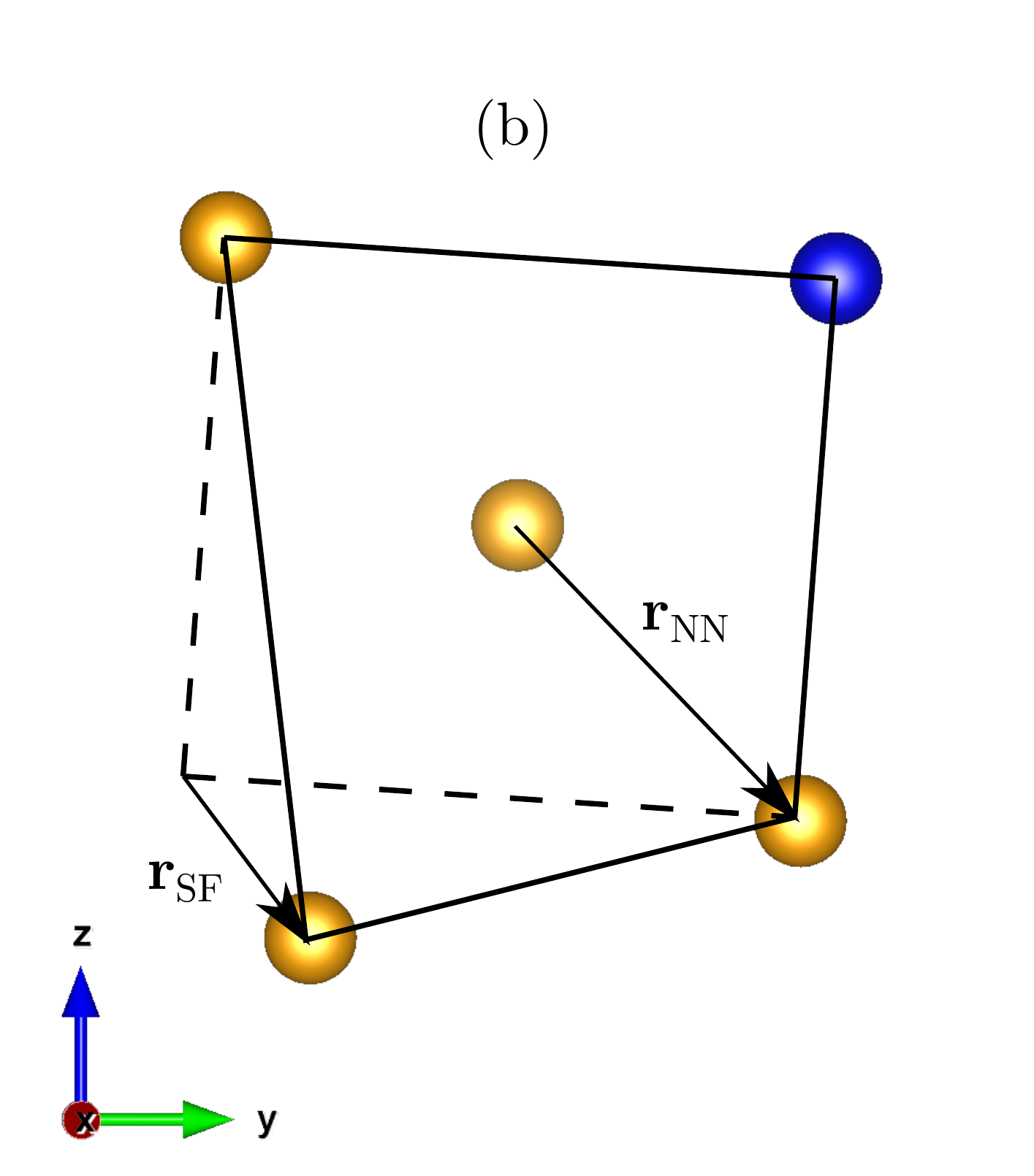}
    \caption{(a) [100] view of Mo$_3$Ta at 15\% uniaxial strain. Tick length units are {\AA}. The size of the atom denotes the height of the atom in the (100) direction, with smaller atoms placed higher. Stacking faults (highlighted using red lines) occur along the [011] plane, denoted by black dashed lines. (b) A small cluster of atoms at the origin (represented by solid lines in (a)) illustrates the displacement vector $\bm{r}_{\mathrm{SF}}$ associated with the dislocation. For this system, $\vert \bm{r}_{\mathrm{SF}} \vert = 0.355\vert \bm{r}_{\mathrm{NN}}\vert$, where $\vert \bm{r}_{\mathrm{NN}} \vert$ is the nearest neighbor separation.}
    \label{fig:stacking-fault-Mo3Ta1}
\end{figure}

\begin{figure}
    \centering
    \includegraphics[width=\linewidth]{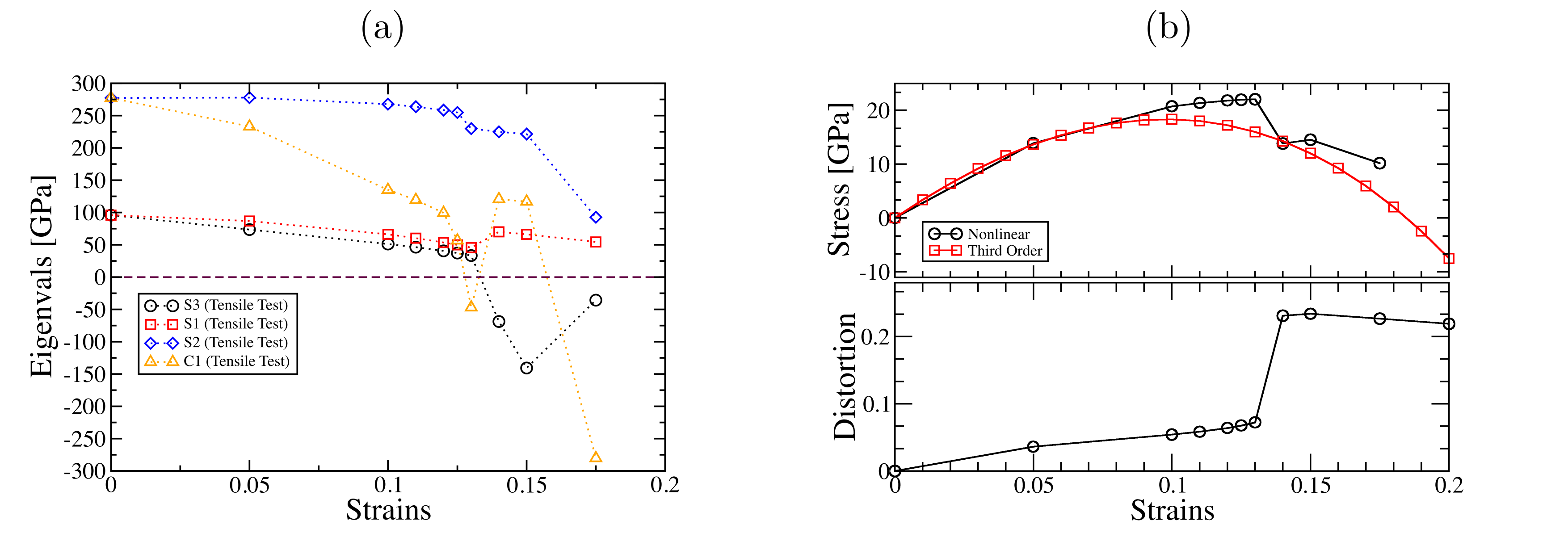}
    \caption{(a) Wallace tensor eigenvalues of Mo$_3$Ta at different uniaxial strains. (b) The stress and lattice distortion (in units of \AA/atom) of Mo$_3$Ta obtained from DFT at different uniaxial strain values.}
    \label{fig:Mo3Ta1-distortion}
\end{figure}

Stacking faults were observed in several alloys (see Figure \ref{fig:stacking-fault-Mo3Ta1} and supplementary material \cite{supplementary}).The transformation impacts the Wallace tensor 
eigenvalues (see Figure \ref{fig:Mo3Ta1-distortion}(a))
Examine the eigenvalues for Mo$_3$Ta, shown in Figure \ref{fig:Mo3Ta1-distortion}(a). Eigenvectors S1,S2 and S3 represent shear modes while C1 represents the cleavage/extensional mode. An additional extensional mode (C2) is also present, however it is not shown in Figure \ref{fig:Mo3Ta1-distortion}(a) since it is an order of magnitude larger than the others and does not affect stability. Extensional failure occurs at 13\%. With further increase in strain, the system finds an alternate solution with shear failure instead of extensional - this is the stacking fault. It is unclear how this can be interpreted in terms of ductility. Shear failure is known to be an indicator of intrinsic ductility, but in this case, it is occurring \textit{after} extensional failure, which has been previously interpreted as fracture. The effect of the stacking fault can also be seen in the stress-strain and lattice distortion curves for Mo$_3$Ta, shown in Figure \ref{fig:Mo3Ta1-distortion}(b). The stacking fault relieves some of the stress present in the system. It also causes a sharp increase in the lattice distortion, which is expected since the system has transformed.

\subsection{Slip}
\begin{figure}
    \centering
    \includegraphics[width=\linewidth]{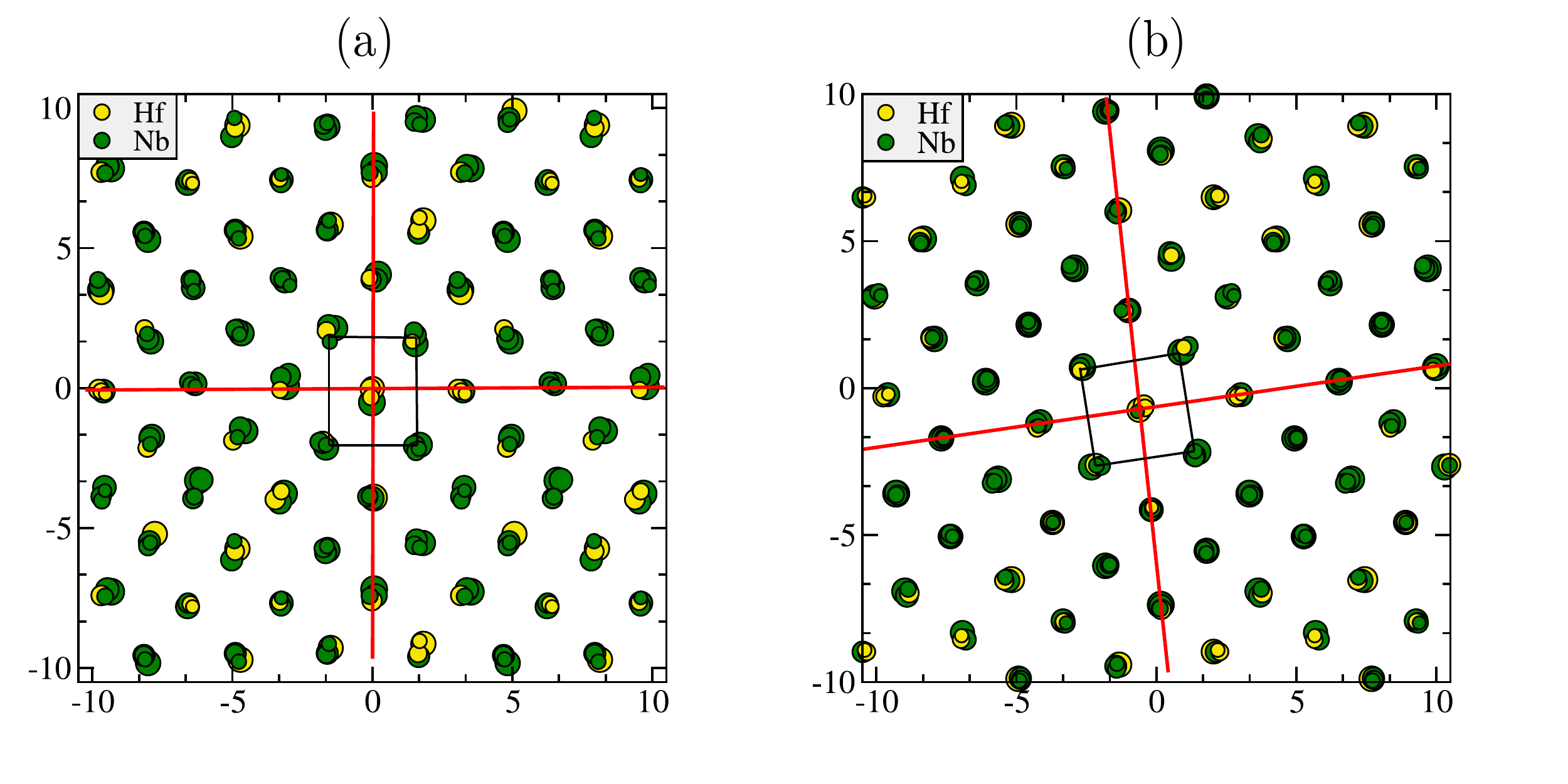}
    \caption{Nb$_3$Hf at (a) 12.5\% uniaxial strain and (b) 15\% uniaxial strain as seen along [100] direction. Tick length units are in {\AA}. The solid red lines highlight the rotation. Solid black lines represent a small cluster of atoms near the origin.}
    \label{fig:Nb3Hf1-dislocation}
\end{figure}

\begin{figure}
    \centering
    \includegraphics[width=\linewidth]{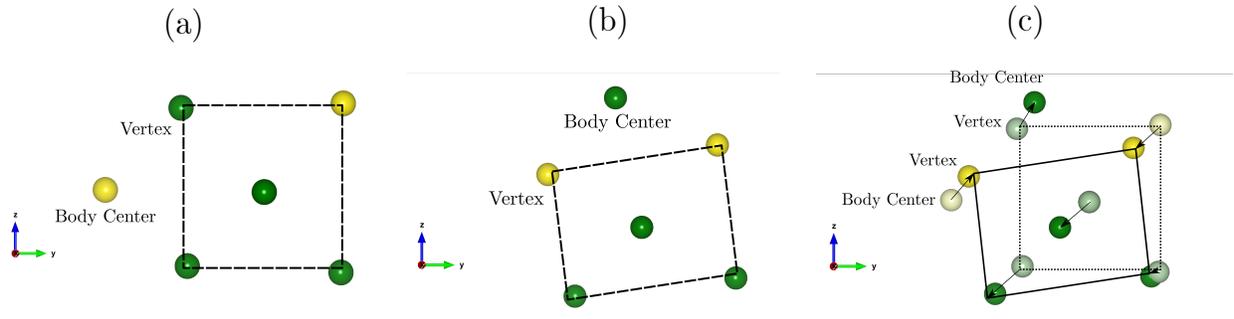}
    \caption{Cluster of atoms taken from Figure \ref{fig:Nb3Hf1-dislocation} consisting of a central atom near the origin surrounded by near neighbors, for Nb$_3$Hf at (a) 0\% (b) 15\% uniaxial strain. Both the clusters are superimposed in (c) to demonstrate the transformation. Green atoms represent Nb and gold atoms represent Hf. In (c), lighter atoms are from (a) and darker atoms are from (b).}
    \label{fig:Nb3Hf-neighbor-analysis}
\end{figure}

Compare the Nb$_3$Hf structures at 12.5\% and 15\% strain, shown in Figures \ref{fig:Nb3Hf1-dislocation}(a) and \ref{fig:Nb3Hf1-dislocation}(b). There is a noticeable rotation in the structure at 15\% strain. We can understand this transformation by closely examining the structure. Figure \ref{fig:Nb3Hf-neighbor-analysis} shows one particular Nb atom (located near the origin) and its near neighbors along the (100) direction. Going from 0\% to 15\%, we see that the species distribution of the near neighbors changes. By superimposing the clusters, we see that some of the atoms have slipped along the [011] plane. The bottom left Nb atom, which was initially a ``vertex" atom ends up as a ``body-center" atom. The ``body-center" Hf atom on the left slips and becomes a ``vertex" atom. This slip transformation gives rise to the rotation observed in Figure \ref{fig:Nb3Hf1-dislocation}(a).

\begin{figure}
    \centering
    \includegraphics[width=\linewidth]{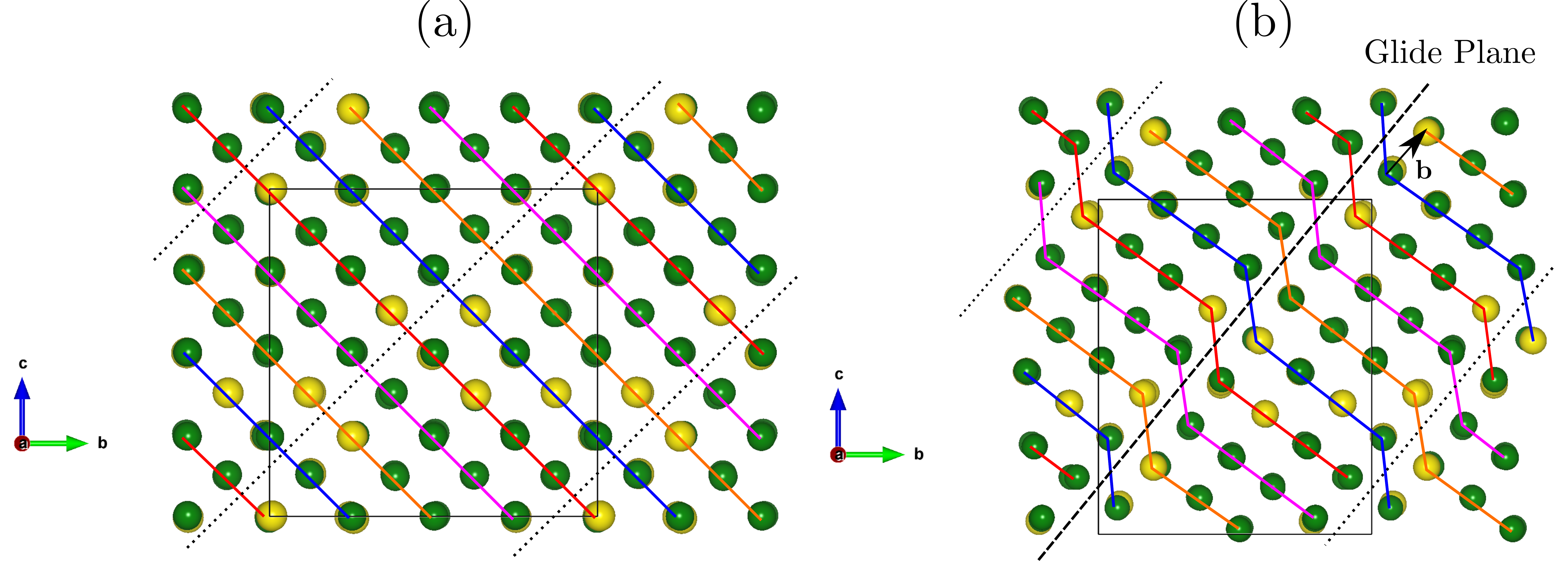}
    \caption{Structure of Nb$_3$Hf at (a) 0\% uniaxial strain (b) 15\% uniaxial strain. The black arrow represents the Burgers vector $\mathbf{b}$. The lattice planes are highlighted using colored lines. The glide plane, which contains both the Burgers vector and the dislocation line is denoted by dashed lines.}
    \label{fig:planes-slip}
\end{figure}

This effect is observed throughout the system. Figure \ref{fig:planes-slip} shows a zoomed out view at 0\% and 15\% strain. Here we can draw the lattice planes along the same atoms in both cases. This allows us to easily view the slip plane. The Burgers vector $b$ is along the [011] direction and (0$\bar{1}$1) is the glide plane which contains both the Burgers vector and the dislocation line. Since our calculations only provide the structure after the dislocation has fully passed, we cannot determine the line of dislocation. 

\begin{figure}
    \centering
    \includegraphics[width=\linewidth]{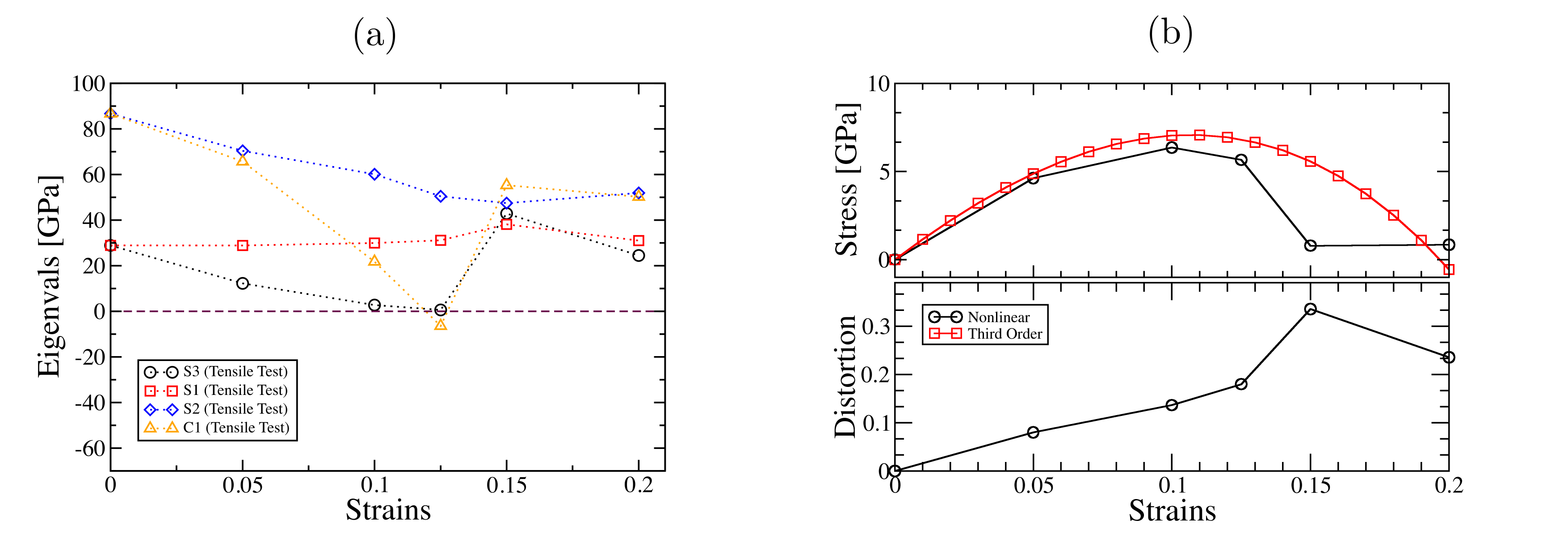}
    \caption{(a) Wallace tensor eigenvalues of Nb$_3$Hf at different uniaxial strains. (b) The stress and lattice distortion (in units of \AA/atom) of Nb$_3$Hf at different uniaxial strain values.}
    \label{fig:Nb3Hf1-distortion}
\end{figure}

The effect of this transformation on elastic stability can be observed from the evolution of the Wallace tensor eigenvalues. After C1 goes negative at 12.5\%, there is a large upward spike in both C1 and S3 eigenvalues at 15\% strain, restoring mechanical stablility. This transformation results in a large drop in stress as seen in Figure \ref{fig:Nb3Hf1-distortion}(b). However, it is not clear if this effect helps provides ductility, since it occurs after extensional failure.

\subsection{Twinning}
\begin{figure}
    \centering
    \includegraphics[width=\linewidth]{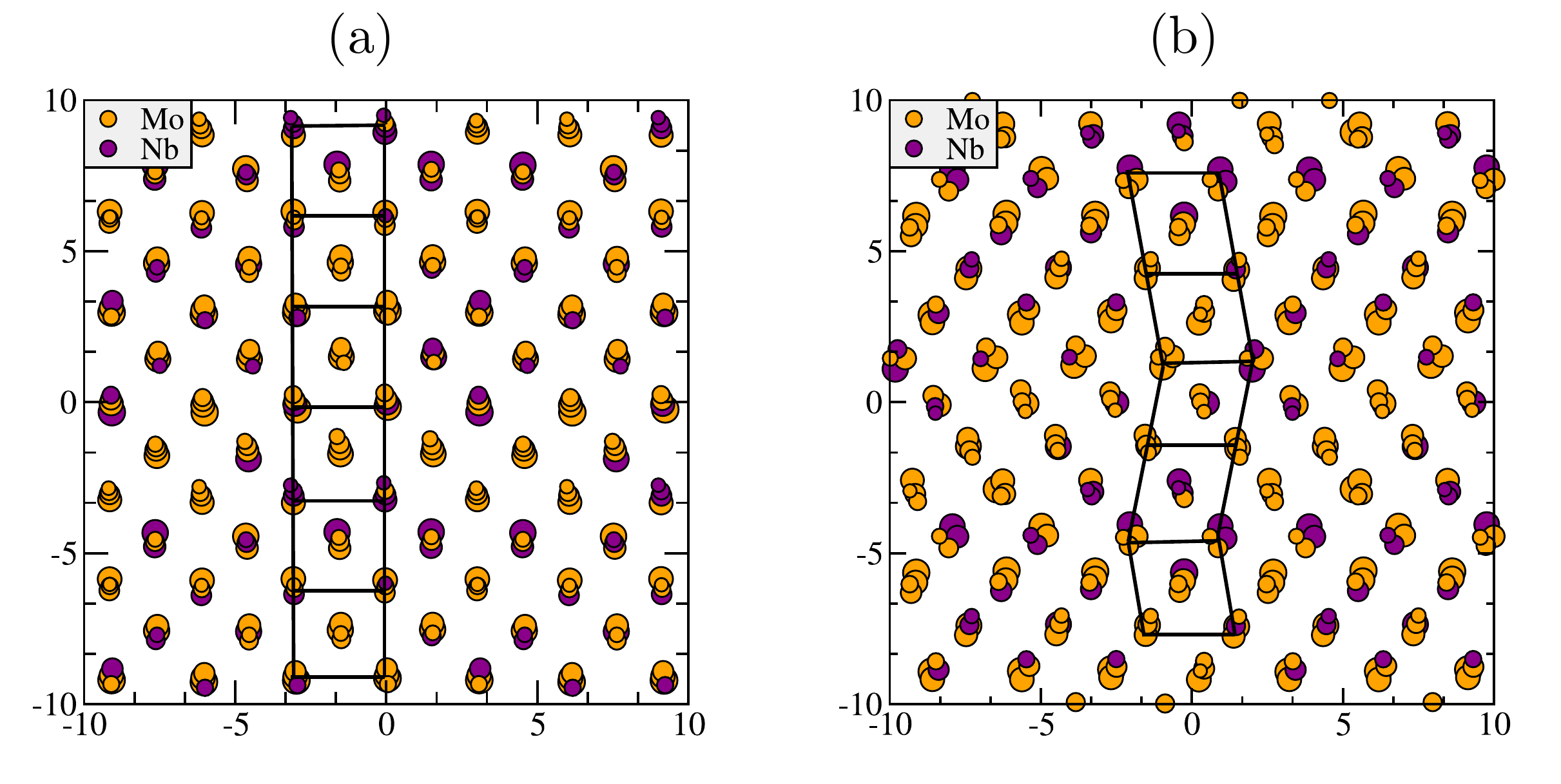}
    \caption{Mo$_3$Nb at (a) 15\% uniaxial strain as seen along [001] direction. The black lines denote the conventional BCC cell. (b) 20\% uniaxial strain as seen along [001] direction. The black lines denote the transformed BCC cell, resulting in twinning patterns.}
    \label{fig:Mo3Nb-twinning}
\end{figure}
Twinning is clearly seen for the Mo$_3$Nb system along the (001) direction (Figure \ref{fig:Mo3Nb-twinning}(a) and (b)). The square BCC cells, denoted by solid black lines, become rhombus-shaped. This phenomenon is referred to as the splay transformation, and transforms tetragonal structure to orthorhombic. It has been observed in previous computational studies of BCC refractory metals \cite{qi,splay}. This is a shear effect and occurs in the plane perpendicular to the stretching direction. In pure Nb, the splay transformation ensures mechanical stability of the crystal and is a mechanism for intrinsic ductility.

\begin{figure}
    \centering
    \includegraphics[width=\linewidth]{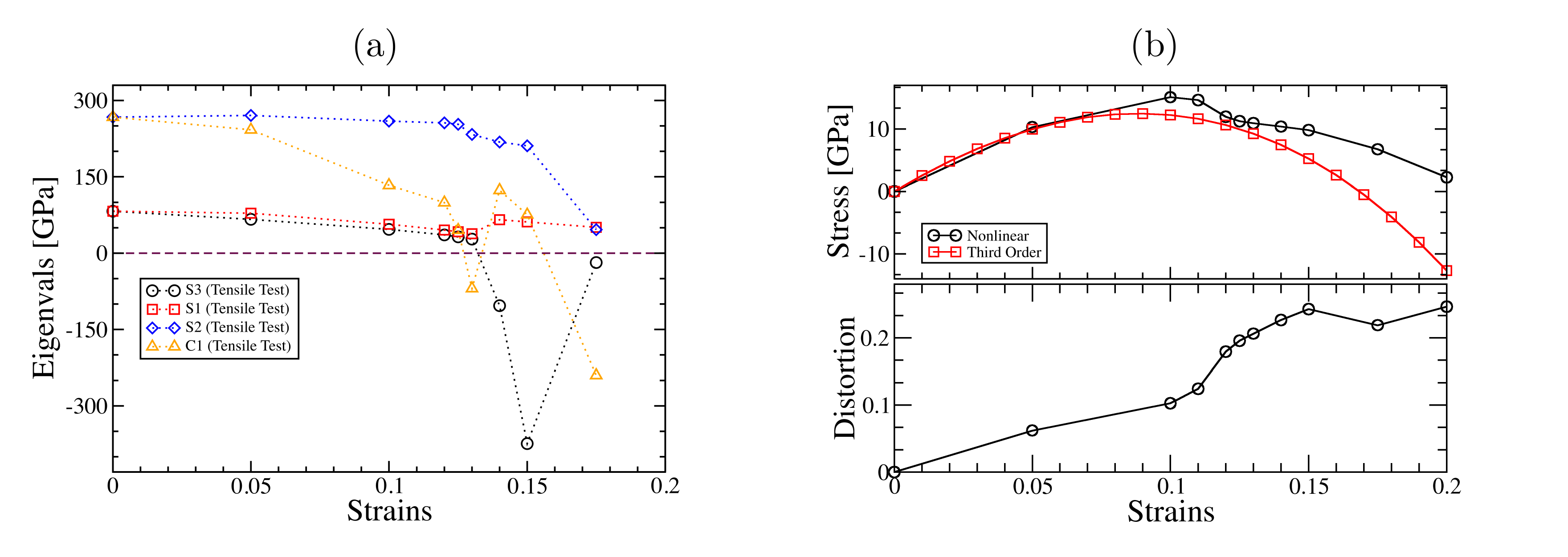}
    \caption{(a) Wallace tensor eigenvalues for Mo$_3$Nb.  Extensional failure occurs before the structure twins. (b) The stress and lattice distortion (in units of \AA/atom) of Nb$_3$Hf at different uniaxial strain values.}
    \label{fig:Mo3Nb-eigenvalues}
\end{figure}

Figure \ref{fig:Mo3Nb-eigenvalues} shows the eigenvalues for Mo$_3$Nb. Twinning is first observed at 17.5\% strain, which is wellbeyond the critical strain for extensional failure. This implies that twinning does not help preserve the mechanical stability of this system and most likely does not play a role in the ductility of the alloy. For all the alloys studied in this work, the onset of twinning occurs after extensional failure. Further data on eigenvalues and twinning critical strains for all the systems that show this effect is present in the supplementary material \cite{supplementary}. 
\subsection{No transformation}
Table \ref{tab:no-transformation} lists the alloys which show steadily increasing lattice distortion without stacking faults, slip or twinning. The maximum stress observed for most of these systems are low in comparison to the other alloys. The supplementary material contains stress, lattice distortion and eigenvalue plots for these systems \cite{supplementary}.
\begin{table}[]
\begin{tabular}{|c|c|c|}
\hline
         & \multicolumn{1}{l|}{Strain at Stress Maximum} & Stress Maximum (GPa) \\ \hline
HfNbTiZr & 0.125                                         & 2.25           \\ \hline
Nb$_3$Si    & 0.125                                         & 3.16           \\ \hline
NbMoTiSi & 0.11                                          & 3.54           \\ \hline
Nb$_3$Al    & 0.10                                          & 4.08           \\ \hline
NbMoTiAl & 0.14                                          & 4.28           \\ \hline
Mo$_3$Si    & {\color[HTML]{000000} 0.05}                   & 5.47           \\ \hline
Mo$_3$Ru    & {\color[HTML]{000000} 0.10}                   & 6.36           \\ \hline
Mo$_3$Al    & 0.11                                          & 9.59           \\ \hline
\end{tabular}
\caption{Table listing the alloys which did not show any stacking faults, twinning or slip.}
\label{tab:no-transformation}
\end{table}

\section{Analysis}
\subsection{Comparing existing parameters}
\begin{figure}
\centering
\includegraphics[width=\linewidth]{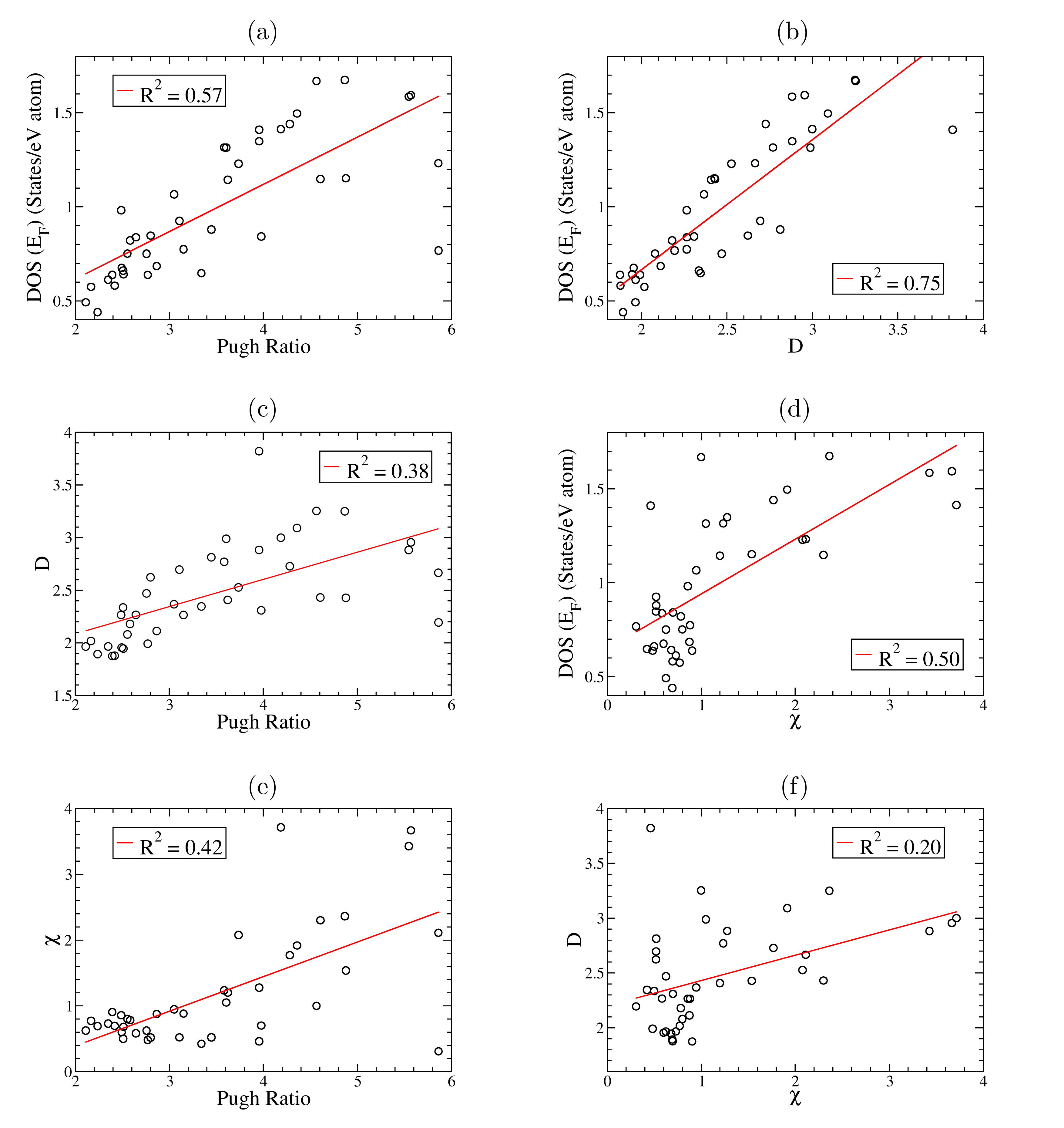}
\caption{Scatter plots (a)-(f) with regression lines, comparing the existing ductility parameters D, $\chi$, Pugh Ratio and the density of states at the Fermi energy. The R$^2$ values for the regression are shown in the legends of each plot.}
\label{fig:existing-parameter-comparison}
\end{figure}

The Pugh Ratio $K/G$ (as defined in Pugh's paper \cite{pugh}), D parameter, $\chi$ parameter and the density of states at the Fermi level have all been previously used as measures of intrinsic ductility \cite{tu-gsfe,surrogate,winter,pugh-application,glass-application-1,glass-application-2}. It is therefore useful to check the consistency of these parameters. Figures \ref{fig:existing-parameter-comparison}(a)-(f) show the correlations among the six possible pairs of four parameters calculated for all the binaries and quarternary systems. In most of these cases, a proportionality can be observed. A high D-parameter corresponds to a high Pugh-ratio, high density of states at the Fermi level and high $\chi$-parameter. This is consistent, as for each of these parameters a high value implies ductility and low value implies brittle failure. The scatter in the plots reflects the widely differing methodologies used for each parameter that capture different aspects of properties linked to ductility. 

\subsection{Peak stress as a ductility parameter}
\begin{figure}
\centering
\includegraphics[width=\linewidth]{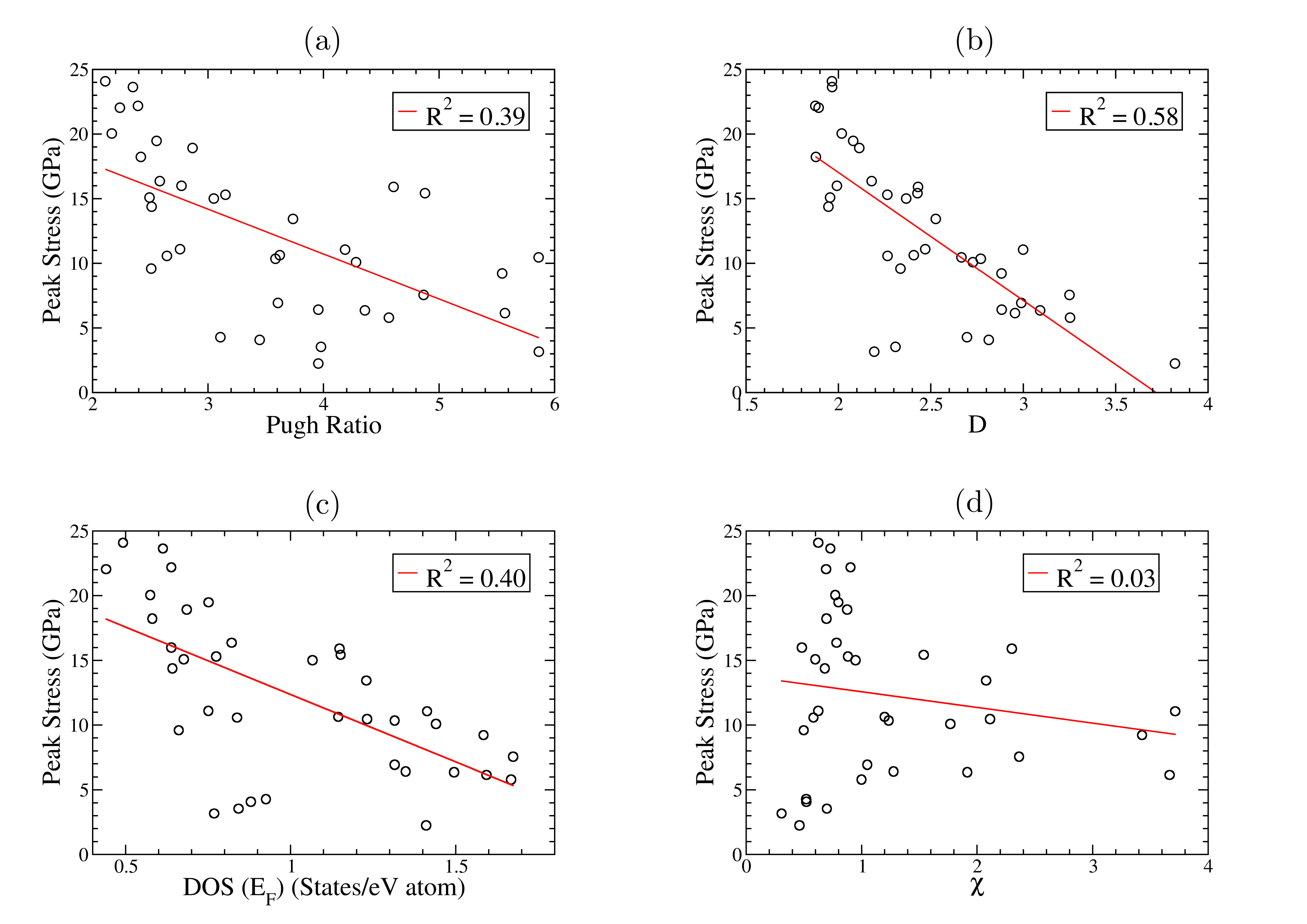}
\caption{Scatter plots (a)-(d) with regression lines, comparing the peak stress obtained from \textit{ab}-\textit{initio} tensile test with the existing ductility parameters D, $\chi$, Pugh Ratio and the density of states at the Fermi energy. The R$^2$ values for the regression are shown in the legends of each plot.}
\label{fig:stress-max-comparison}
\end{figure}

In addition to capturing stacking fault, slip and twinning patterns, the peak stress, defined as the maximum stress obtained in the \textit{ab}-\textit{initio} tensile test, functions as an intrinsic ductility parameter. To see this, we plot the tensile test peak stress against the four existing ductility parameters for all the binaries and quarternaries (Figure \ref{fig:stress-max-comparison}(a)-(d)). With the exception of $\chi$, there is an inverse proportionality between the peak stress and the existing ductility parameters. This trend allows us to establish the tensile test peak stress as a ductility parameter with the inverse criteria for ductility. A high peak stress would likely cause brittle failure, while a low peak stress would enhance ductility. The $\chi$ parameter plot is an outlier; as with the previous case, the varying methodologies used for the parameters determine their consistency with other parameters.
\subsection{Valence electron count}
\begin{figure}
\centering
\includegraphics[width=\linewidth]{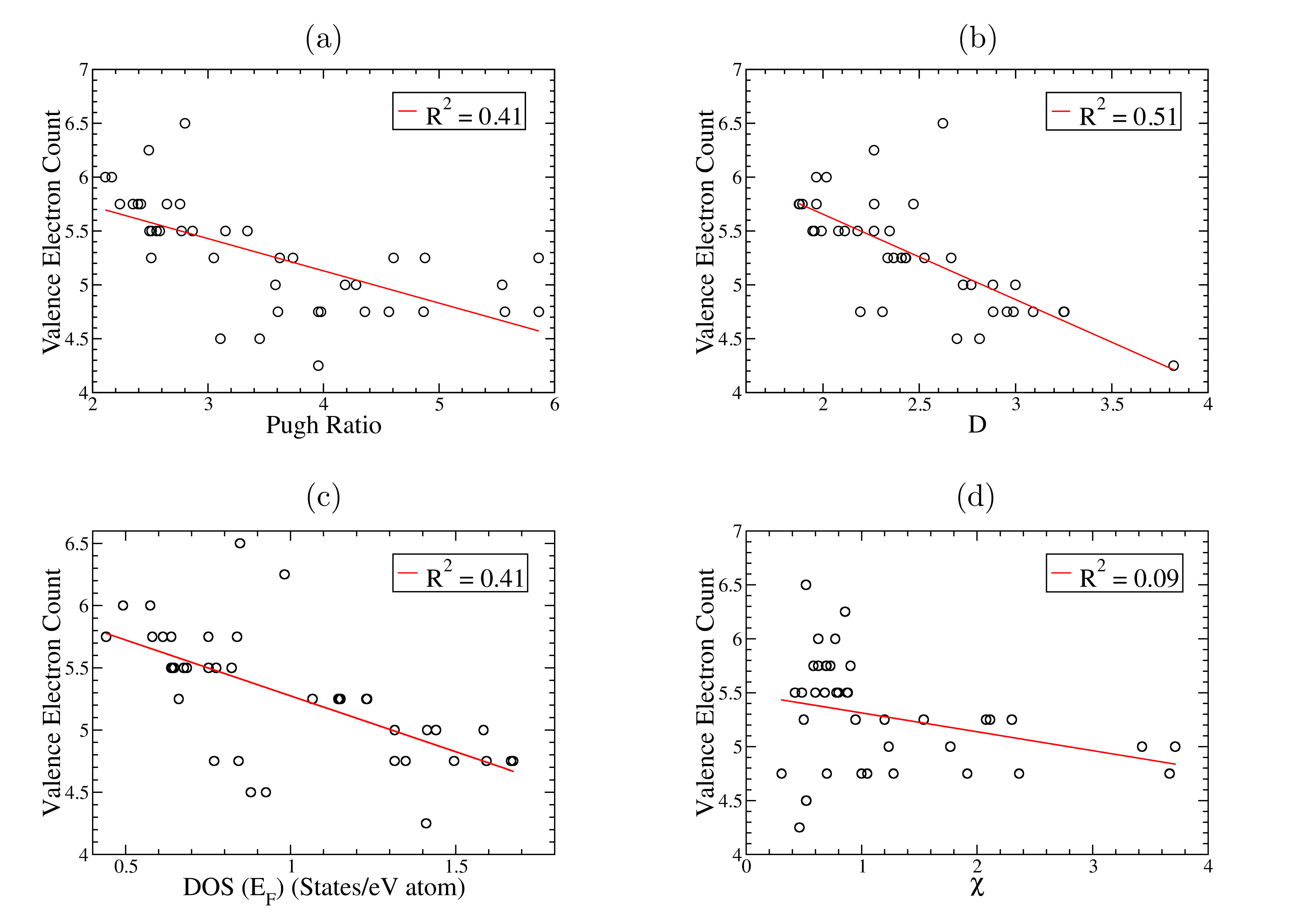}
\caption{Scatter plots (a)-(d) with regression lines, comparing the VEC with the existing ductility parameters D, $\chi$, Pugh Ratio and the density of states at the Fermi energy. The R$^2$ values for the regression are shown in the legends of each plot.}
\label{fig:vec-existing-comparison}
\end{figure}

\begin{figure}
\centering
\includegraphics[width=0.5\linewidth]{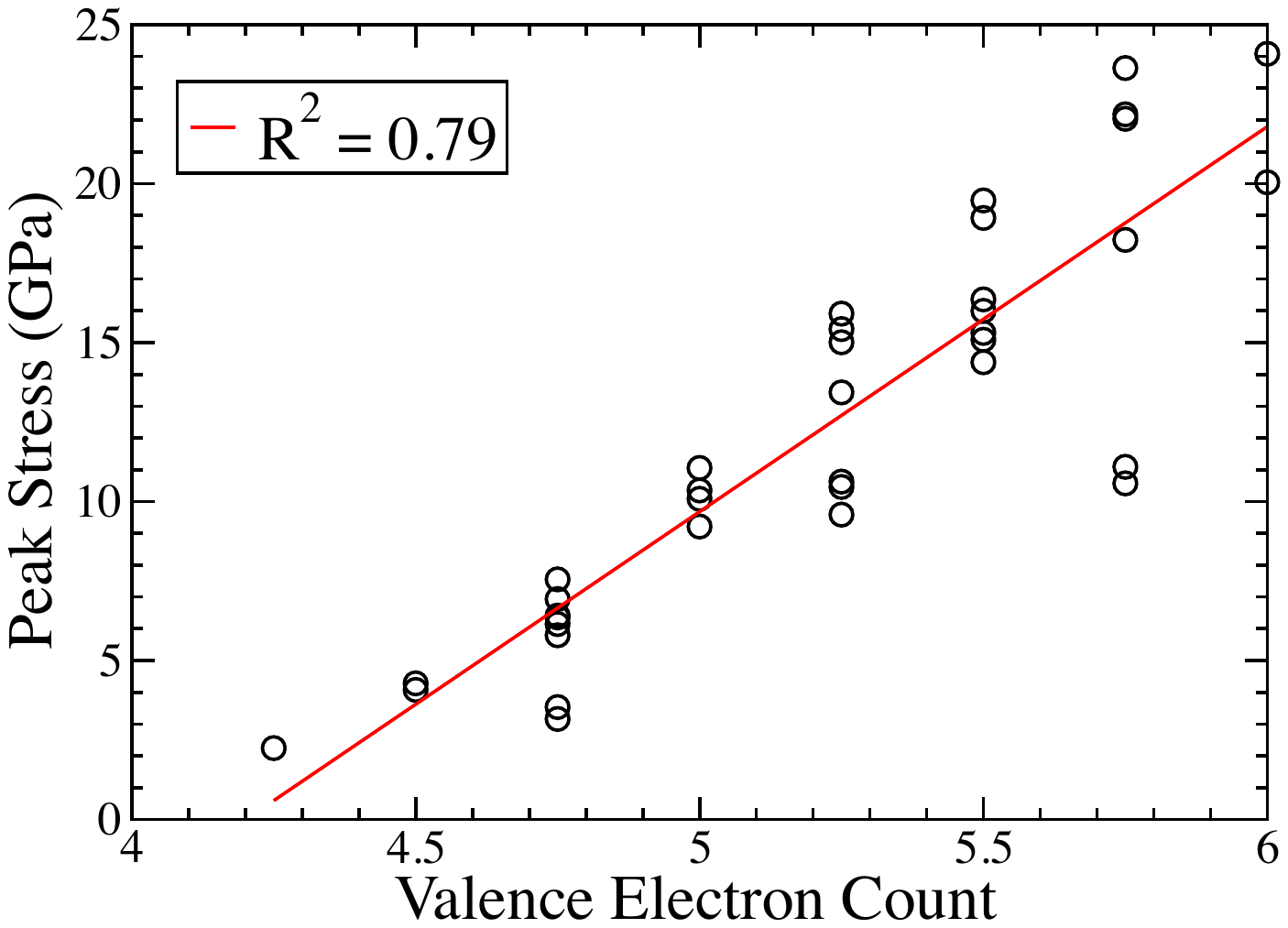}
\caption{Scatter plot with regression lines, comparing the VEC with the tensile test peak stress. The R$^2$ values for the regression are shown in the legend.}
\label{fig:vec-peak-stress}
\end{figure}

We compare the valence electron count (VEC) for all the system against the five ductility parameters Pugh ratio, D parameter, $\chi$ parameter, density of states at Fermi level and the tensile test peak stress (Figures \ref{fig:vec-existing-comparison}(a)-(d) and \ref{fig:vec-peak-stress}). The VEC linearly decreases with the four existing parameter and is directly proportional to the peak stress. This implies that the VEC could potentially be used as a ductility parameter - high VEC would imply brittleness while low VEC would imply ductility. This is a useful observation since the VEC can be obtained by simply looking at the composition and hence no calculations are required. Since many different compositions could have the same VEC, other ductility parameters will be needed for a more precise assessment. However sorting by VEC could be a helpful preliminary step in ductile alloy design.
\section{Discussion}
\subsection{Workflow for the \textit{ab}-\textit{initio} tensile test}
\begin{figure}
    \centering
    \includegraphics[width=0.6\linewidth]{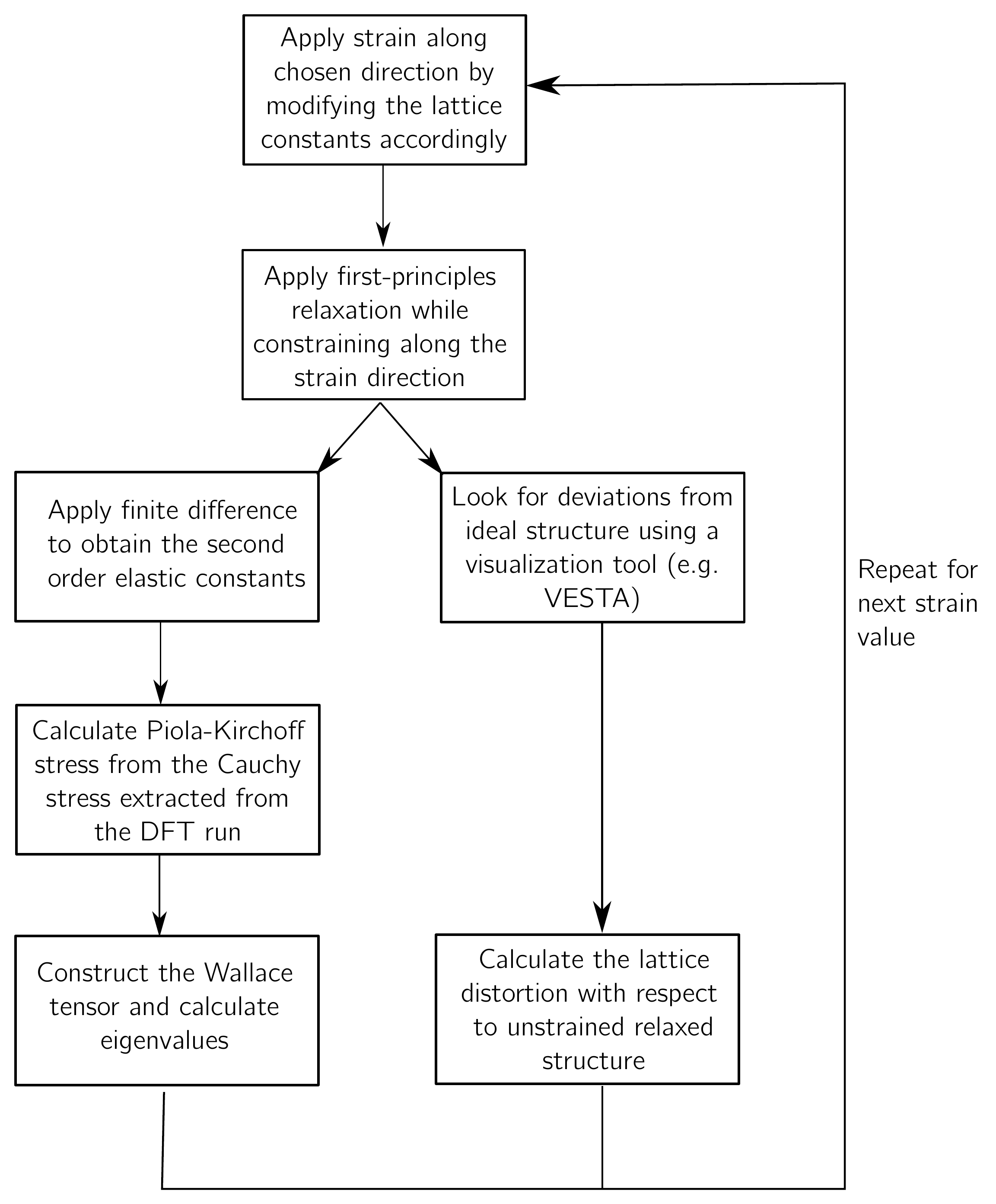}
    \caption{Chart illustrating the workflow for the $ab$-$initio$ tensile test method.}
    \label{fig:workflow}
\end{figure}

To promote the ease of use, we provide a detailed workflow for the $ab$-$initio$ tensile test in Figure \ref{fig:workflow}. Broadly there are two major parts in this method - careful analysis of the relaxed structures and the calculation of Wallace tensor eigenvalues at different strains. We use simple python scripts for most of the post-processing steps, like  extracting elastic constants, stress, calculating eigenvectors etc. All these scripts used for the post-processing have been made available on github. \cite{Elastictoolkit}. Computational details for the first principles calculations are available in the supplementary material \cite{supplementary}.
\subsection{Comments on Ductility Parameters}

However, there are drawbacks to using any single parameter to characterize intrinsic ductility. Every intrinsic ductility parameter has underlying assumptions that cause it to deviate from experiment. The Pugh Ratio and $\chi$ parameters assume ductility can be described by zero-strain second order elastic constants. The D-parameter assumes ductility arises from crack tip blunting, which is only one of several possible mechanisms that could provide ductility. The tensile test approach also relies on elastic constants, albeit at finite strain. Since each model parameter is idealized in different ways, precise agreement between the parameters is not expected. We can see this from the scatter plots of Figures \ref{fig:existing-parameter-comparison},\ref{fig:stress-max-comparison},\ref{fig:vec-existing-comparison} and \ref{fig:vec-peak-stress}. There are useful general trends which we discussed in the previous section, but there is also significant variance from the least squares fit. These parameters are useful when applied to a large number of systems in order to identify a group of potentially ductile compositions. But when analyzing specific alloys, these parameters provide conflicting assessments of the ductility. Fine-grained analysis of ductility requires realistic simulations which include defects and grain boundaries and account for all the possible ductility providing mechanisms that could arise in the system. The $ab$-$initio$ tensile test is a small step in that direction, since it produces realistic effects like slip and twinning that occur at large strains. However, we still use idealized unit cells. Performing the $ab$-$initio$ tensile test on very large unit cells with different grains would be an interesting future project which could produce more exciting results. Soft phonon modes with non-zero wavenumber can contribute to the elastic stability in BCC structures \cite{phonon-instability}. However calculating the phonon spectra for the strained supercell structures is an extremely intensive task, and has been left as a future exercise. Obtaining a material agnostic parameter that can accurately and reliably predict ductility is an extremely complicated proposition and still remains an open problem in the field.

\section{Conclusion}
In this work, we used the $ab$-$initio$ tensile test method to study intrinsic ductility. This method was applied to a large number of Mo and Nb based binaries and various failure mechanisms like twinning, stacking fault and slip were uncovered. The eigenvalues of the symmetric Wallace tensor were also calculated, and non-monotonicities in the eigenvalue-strain curves were correlated with the different failure modes. The peak stress and VEC were linked to ductility by comparing them to the D-parameter, $\chi$-parameter, Pugh ratio and the density of states at the Fermi level, all of which have been used previously as indicators of ductility. The novelty and limitations of this method were discussed, along with the difficulties in accurate ductility estimation. Finally, we provide some ideas for improving the $ab$-$initio$ tensile test approach.

\begin{acknowledgements}
This work was supported by the ARPA-E program at CMU (Grant No. DE-AR0001430) and at NETL. This work was funded by the Department of Energy, ARPA-e, an agency of the United States Government. It was also supported by the Department of Energy Grant No. DE-SC0014506. We acknowledge useful discussions with Amit Samanta (LLNL) on ductility parameter analysis. This research used resources of the National Energy Research Scientific Computing Center (NERSC), a DOE Office of Science User Facility supported by the Office of Science under Contract No. DE-AC02-05CH11231 using NERSC award ALCC-ERCAP0022624.
\end{acknowledgements}

\end{document}